\documentstyle[12pt,epsf,epsfig]{article}
\newcount\timecount
\newcount\hours \newcount\minutes  \newcount\temp \newcount\pmhours

\hours = \time
\divide\hours by 60
\temp = \hours
\multiply\temp by 60
\minutes = \time
\advance\minutes by -\temp
\def\hour{\the\hours}
\def\minute{\ifnum\minutes<10 0\the\minutes
            \else\the\minutes\fi}
\def\clock{
\ifnum\hours=0 12:\minute\ AM
\else\ifnum\hours<12 \hour:\minute\ AM
      \else\ifnum\hours=12 12:\minute\ PM
            \else\ifnum\hours>12
                 \pmhours=\hours
                 \advance\pmhours by -12
                 \the\pmhours:\minute\ PM
                 \fi
            \fi
      \fi
\fi
}

\def\monthname{\relax\ifcase\month 0/\or January\or February\or
   March\or April\or May\or June\or July\or August\or September\or
   October\or November\or December\else\number\month/\fi}

\def\bold#1{\setbox0=\hbox{$#1$}%
     \kern-.025em\copy0\kern-\wd0
     \kern.05em\copy0\kern-\wd0
     \kern-.025em\raise.0433em\box0 }

\def\gappeq{\mathrel{\rlap {\raise.5ex\hbox{$>$}}
{\lower.5ex\hbox{$\sim$}}}}

\def\lappeq{\mathrel{\rlap{\raise.5ex\hbox{$<$}}
{\lower.5ex\hbox{$\sim$}}}}

\def\ga{\mathrel{\raise.3ex\hbox{$>$\kern-.75em\lower1ex\hbox{$\sim$}}}}
\def\la{\mathrel{\raise.3ex\hbox{$<$\kern-.75em\lower1ex\hbox{$\sim$}}}}
\def\gev{{\rm \, Ge\kern-0.125em V}}
\def\tev{{\rm \, Te\kern-0.125em V}}
\def\beq{\begin{equation}}
\def\eeq{\end{equation}}

\def\m12{m_{1\!/2}}

\begin{document}
\begin{titlepage}
\pagestyle{empty}
\baselineskip=21pt
\rightline{hep-ph/0009355}
\rightline{CERN--TH/2000-293, MPI-PhE/2000-23}
\rightline{ACT-14-00, CTP-TAMU-31/00}
\rightline{UMN--TH--1925/00, TPI--MINN--00/49}
\vskip 0.05in
\begin{center}
{\large{\bf
What if the Higgs Boson Weighs 115 GeV?}}
\end{center}
\begin{center}
\vskip 0.05in
{{\bf John Ellis}$^1$, 
{\bf Gerardo Ganis}$^2$,
{\bf D.V. Nanopoulos}$^3$ and
{\bf Keith A.~Olive}$^{1,4}$}\\
\vskip 0.05in
{\it
$^1${TH Division, CERN, Geneva, Switzerland}\\
$^2${Max-Planck-Institut f\"ur Physik, Munich, Germany; \\
present address: INFN, Universit\`a di Roma II `Tor Vergata', Rome,
Italy}\\
$^3${Department of Physics, Texas A \& M University,
College Station, TX~77843, USA; \\
Astroparticle Physics Group, Houston
Advanced Research Center (HARC), \\
Mitchell Campus,
Woodlands, TX~77381, USA; \\
Chair of Theoretical Physics,
Academy of Athens,   
Division of Natural Sciences,
28~Panepistimiou Avenue,
Athens 10679, Greece}\\
$^4${Theoretical Physics Institute, School of Physics and Astronomy,\\
University of Minnesota, Minneapolis, MN 55455, USA}\\
}
\vskip 0.05in
{\bf Abstract}
\end{center}
\baselineskip=18pt \noindent

If the Higgs boson indeed weighs about 114 to 115 GeV, there must be new
physics
beyond the Standard Model at some scale $\la 10^6$~GeV. The most plausible
new physics is supersymmetry, which predicts a Higgs boson weighing $\la
130$~GeV. In the CMSSM with $R$ and CP conservation, the existence,
production and detection of a 114 or 115~GeV Higgs boson is possible if
$\tan\beta \ga 3$. 
However, for the radiatively-corrected Higgs
mass to be this large, sparticles should be relatively heavy: $m_{1/2} \ga
250$~GeV, probably not detectable at the Tevatron collider and perhaps not
at a low-energy $e^+ e^-$ linear collider. In much of the remaining CMSSM
parameter space, neutralino-$\tilde \tau$ coannihilation is important for
calculating the relic neutralino density, and we explore implications for
the elastic neutralino-nucleon scattering cross section.

\end{titlepage}
\baselineskip=18pt

At the time of writing, the LEP experiments are not yet able 
to exclude the possibility that the Higgs boson might weigh
about 114 to 115~GeV, and there are several candidate events~\cite{LEPC}
for its production in association with a $Z$ boson~\cite{ZH}, that may
be appearing above the Standard Model background. It is hoped
that the high-energy LEP luminosity used for the
presentations~\cite{LEPC} may be increased substantially
before the accelerator is closed by the end of this year,
enabling the possible signal to be either strengthened or diluted
significantly.
However, it is unlikely that LEP will be able to answer definitively
the question whether there exists a Higgs boson weighing about 114 to 
115~GeV.
Indeed, a definitive answer may not be available for several years,
until either the Fermilab Tevatron collider accumulates enough
luminosity~\cite{Tevatron} and/or the LHC starts up~\cite{LHC}.

Even in these circumstances, it is tempting to speculate on the
interpretation of a possible discovery of
a Higgs boson weighing around 114 to 115~GeV. This might even serve the
useful
purpose of suggesting other signatures that could be correlated
with the existence of such a Higgs boson, whose appearance (absence)
might help to confirm (cast doubt upon) any evidence for its existence.

The first clear statement that can be made is that {\it if} the Higgs
boson weighs about 114 to 115~GeV, {\it there must be new physics at an
energy
scale $\ll M_P$}. This is because of the renormalization of the
effective Higgs potential by the Higgs-top interaction
$\lambda_t {\bar t} t \phi$ (related to the known value of $m_t$) and the
Higgs self-interaction $\lambda
\phi^4$ (related to the putative value of $m_H$). It is well known that,
if either $\lambda_t$
and/or $\lambda$ is too large, the renormalization-group equations
(RGEs) may cause the couplings to blow up, becoming non-perturbative or
even infinite at some energy scale below $M_P$~\cite{toobig}.
Alternatively,
the desired electroweak vacuum may become unstable if $\lambda$ is too
small, since $\lambda_t$ tends to drive the effective Higgs potential
$V(\phi)$ {\it negative} at large $|\phi|$~\cite{toosmall,HR}.
Self-renormalization by
$\lambda$ tries to counteract this effect of $\lambda_t$, but is
overcome if $\lambda$ is too small. Requiring the absence of a second,
undesirable minimum of the effective Higgs potential for any value
$|\phi| \le \Lambda$ therefore provides a lower limit on $\lambda$,
and hence $m_H$, that depends on $\Lambda, m_t$ and (via higher orders
in the RGEs) the strong gauge coupling $\alpha_s$. Conversely, given
$m_H$, and hence $\lambda$,
one has an upper limit on the scale $\Lambda$ up to which the
Standard Model Higgs potential may remain stable, which depends
relatively on the precise values of $m_t$ and $\alpha_s$.
If indeed $m_H = 115$~GeV, one finds that~\cite{toosmall}
\begin{equation}
\Lambda \,\, \la \,\, 10^6 \, {\rm GeV}
\label{upper}
\end{equation}
for the default values $m_t = 175$~GeV and $\alpha_s (m_Z) =
0.118$~\footnote{The
upper limit (\ref{upper}) increases (decreases) by about an order of
magnitude
for $m_t = 170 (180)$~GeV, while being less sensitive to $\alpha_s (m_Z)$
in the range 0.115 - 0.121.}.
Therefore, {\it there must be new physics at some scale $\la 
10^6$}~GeV that averts this instability of the Standard Model Higgs
potential.

It has often been suggested that new physics should be
expected at some scale $\la 1$~TeV, in order to stabilize the gauge
hierarchy. Prominent among early suggestions was that of technicolour,
new strong interactions that would generate a
composite scalar particle weighing about 1~TeV~\cite{TC}. One generally
expects composite models to predict Higgs bosons
much heavier than whatever LEP might be seeing, and technicolour bears
this out. Technicolour models
generally also include light pseudoscalar particles, but these would not
be produced copiously in association with a $Z$ boson~\cite{hole}.
Another class of composite Higgs models invokes ${\bar t}t$ condensation,
but these models also predict~\cite{topcolour} a Higgs boson
that would be heavier than what LEP might be seeing.
In the absence of any viable composite Higgs model, we pursue the
hypothesis that the Higgs is elementary, as generally expected for
small $m_h$, in which case
the most plausible new TeV-scale physics is supersymmetry~\cite{MSSM}.

Circumstantial evidence for supersymmetry around this scale has already
been provided by the possible grand unification of the gauge
couplings, which works fine if sparticles weighing around 1~TeV are
included in their RGEs~\cite{susyGUT}. Moreover, the minimal
supersymmetric extension of
the Standard Model (MSSM) predicts the existence of at least one neutral
Higgs boson weighing $\la 130$~GeV~\cite{sHiggs,HHH}, perfectly consistent
with the possible
direct LEP observation~\cite{LEPC} and with indirect indications from
precision
electroweak data of a relatively light Higgs boson~\cite{precision}:
$m_h = 62^{+53}_{-30}$~GeV, with the one-sided 95\% confidence-level
upper limit $m_h < 170$~GeV~\footnote{The central value may be increased
by $\sim 30$~GeV if new data from BES are used to evaluate
$\alpha_{em}(M_Z)$~\cite{precision}: see also~\cite{Martin}.}. Therefore,
in the
rest of this paper we concentrate on supersymmetric interpretations of the
possible LEP observation of a Higgs boson weighing 114 to 115~GeV. We
assume the
conservation of $R$ parity, so that the lightest neutralino $\chi$ may
constitute the cold dark matter postulated by astrophysicists and
cosmologists~\cite{EHNOS}. We assume also CP conservation for the
tree-level MSSM
parameters, simplifying calculations of the Higgs masses~\cite{CEPW} and
dark matter properties~\cite{CPDM}. 

As we show in this paper, the possible observation of a Higgs boson
weighing 114 to 115~GeV would constrain significantly the sparticle
spectrum in such models~\footnote{We phrase our discussion optimistically
in terms of the observation of such a Higgs boson: our lower limits on 
the sparticle spectrum also apply if LEP only establishes a lower limit
$m_h \ge 114$~GeV.}, and hence the prospects for sparticle detection.
The principal uncertainty in predicting the sparticle mass spectrum is
due to the lack of precision in the measurement of $m_t$, which is
also manifest in our discussion of the
potential of an $e^+ e^-$ linear collider to discover supersymmetry.

We assume a minimal supergravity-inspired model of soft supersymmetry
breaking, namely the constrained MSSM (CMSSM) or minimal supergravity
(mSUGRA), in which universal gaugino masses $m_{1/2}$, scalar masses $m_0$
(including those of the Higgs multiplets) and trilinear supersymmetry
breaking parameters $A$ are input at the supersymmetric grand unification
scale \footnote{It would be an interesting exercise to make a similar
analysis in the context of gauge- or anomaly-mediated models~\cite{bpz},
but this
lies beyond the scope of our work.}. In this framework, the Higgs mixing
parameter $\mu$ can be derived from the other MSSM parameters by imposing
the electroweak vacuum conditions for any given value of $\tan \beta$.

\begin{figure}[htbp]
\begin{center}
\mbox{\epsfig{file=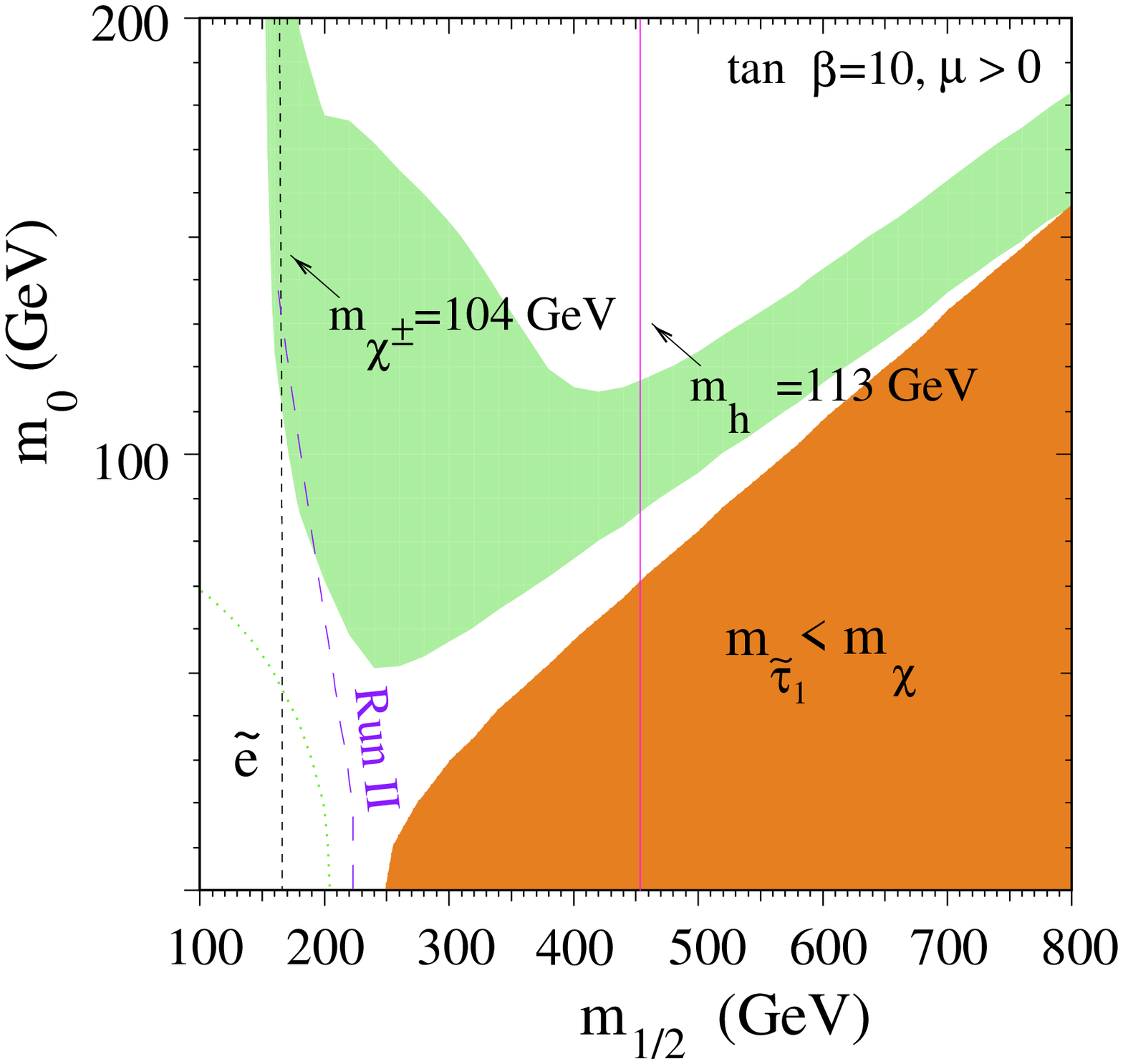,height=8cm}}
\mbox{\epsfig{file=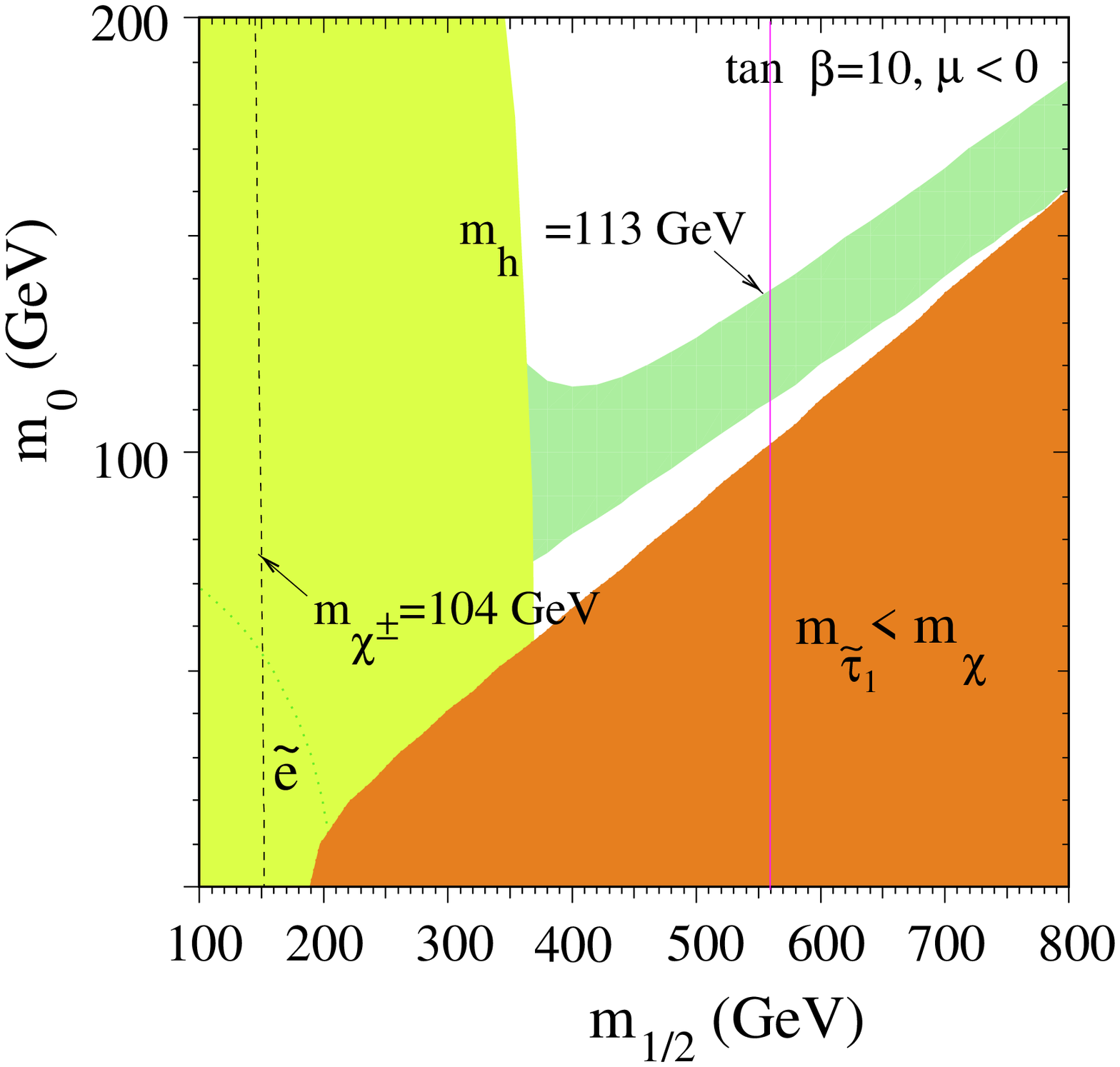,height=8cm}}
\end{center}
\caption[.]{\label{fig:efgo}\it 
The $m_{1/2}, m_0$ plane for the CMSSM with $\tan \beta = 10$, $A = -
m_{1/2}$, and (a) $\mu > 0$, 
(b) $\mu < 0$, showing the region preferred by the cosmological
relic density constraint $0.1 \le \Omega_\chi h^2 \le 0.3$ (medium, green
shading), the
excluded region where $m_{\tilde \tau} < m_\chi$ (dark, brown shading),
and the region disallowed by our $b \rightarrow s \gamma$ analysis
(light shading)~\cite{EFGO}. Also shown as a
near-vertical line is the contour $m_h = 113$~GeV for $m_t =
175$~GeV. For comparison, we also exhibit the reaches of LEP~2 searches
for charginos $\chi^\pm$ and selectrons $\tilde e$, as well as the
estimated reach of the Fermilab
Tevatron collider for sparticle production~\cite{Barger}.}
\end{figure}

Many ingredients in our analysis are apparent from Fig.~\ref{fig:efgo},
including the range of $(m_{1/2}, m_0)$ where the relic neutralino density
is in the range of cosmological interest: $0.1 \la \Omega_{\chi} h^2
\la 0.3$, the excluded region at low $m_0$ and large $m_{1/2}$ where the
lightest
sparticle is a charged ${\tilde \tau}$, and a region at low $m_{1/2}$
for $\mu < 0$ that is excluded~\cite{Gambino} by the experimental value of
$b \rightarrow s \gamma$ decay~\cite{bsgamma}
We recall that the mass of the lightest neutralino $m_\chi
\simeq 0.4 \times m_{1/2}$ over most of the gaugino parameter region of
interest.
As discussed in~\cite{EFGO}, there are important regions of the $(m_{1/2},
m_0)$ plane where the present electroweak vacuum is at best metastable
against
decay into a vacuum where charge and colour are broken (CCB)~\cite{AF}. We
do not
address here the question of the lifetime of the vacuum, which is much
longer than that in the Standard Model for light Higgs mass. However, we
do note that there are regions at large $m_0$ and/or $m_{1/2}$ that are
completely stable against decay into a CCB vacuum~\cite{AF,EFGO}. 

Through radiative corrections~\cite{sHiggs,HHH}, the mass $m_h$ of the
lightest Higgs boson
depends strongly on $m_{1/2}$, but is almost independent of $m_0$, at
least over the range of $m_0$ allowed by the upper limit $\Omega_{\chi}
h^2 \la 0.3$ on the relic neutralino density, as can also be seen in
Fig.~\ref{fig:efgo}.  The Higgs mass $m_h$ also depends significantly on
$m_t$, varying typically by $\pm 3$~GeV, as $m_t$ is varied by $\pm 5$~GeV
around its nominal value $m_t = 175$~GeV.
The uncertainty in $m_t$ carries through to our final bounds on the
sparticle spectrum~\footnote{On the other hand, the
dark matter density calculations are relatively insensitive to $m_t$.},
as discussed later.  There are believed to be similar
uncertainties in $m_h$ associated with the treatment of higher-order QCD
corrections to $m_t$~\cite{HHH}.  Other uncertainties, associated for
example with
higher-order electroweak effects, are believed to be ${\cal O} (1)$ GeV.
We
recall that the preferred range of $m_h$ suggested by LEP is from $114$
to $115$
GeV~\cite{LEPC}.  We derive our lower (upper) limits on the sparticle
spectrum by
finding the values of $m_{1/2}$ required to give $m_h \ge 113
(\le 116)$~GeV for
$m_t = 170, 175$ and $180$ GeV, so as to include some allowance for these
uncertainties. 

\begin{figure}[htb]
\begin{center}
\vskip -3.0in
\mbox{\hskip -.7in \epsfig{file=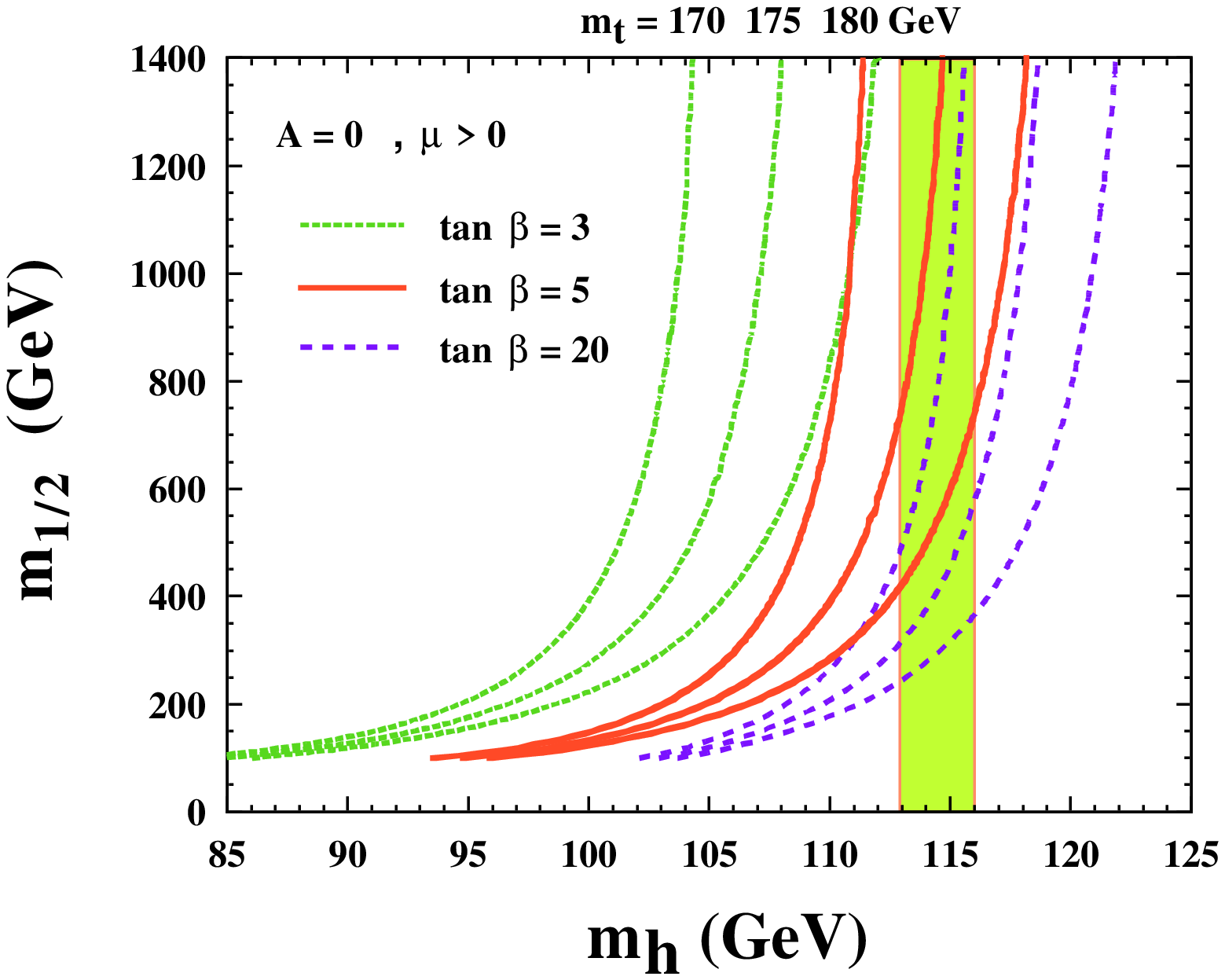,height=15cm}}
\vskip -2.1in
\mbox{\hskip -.7in \epsfig{file=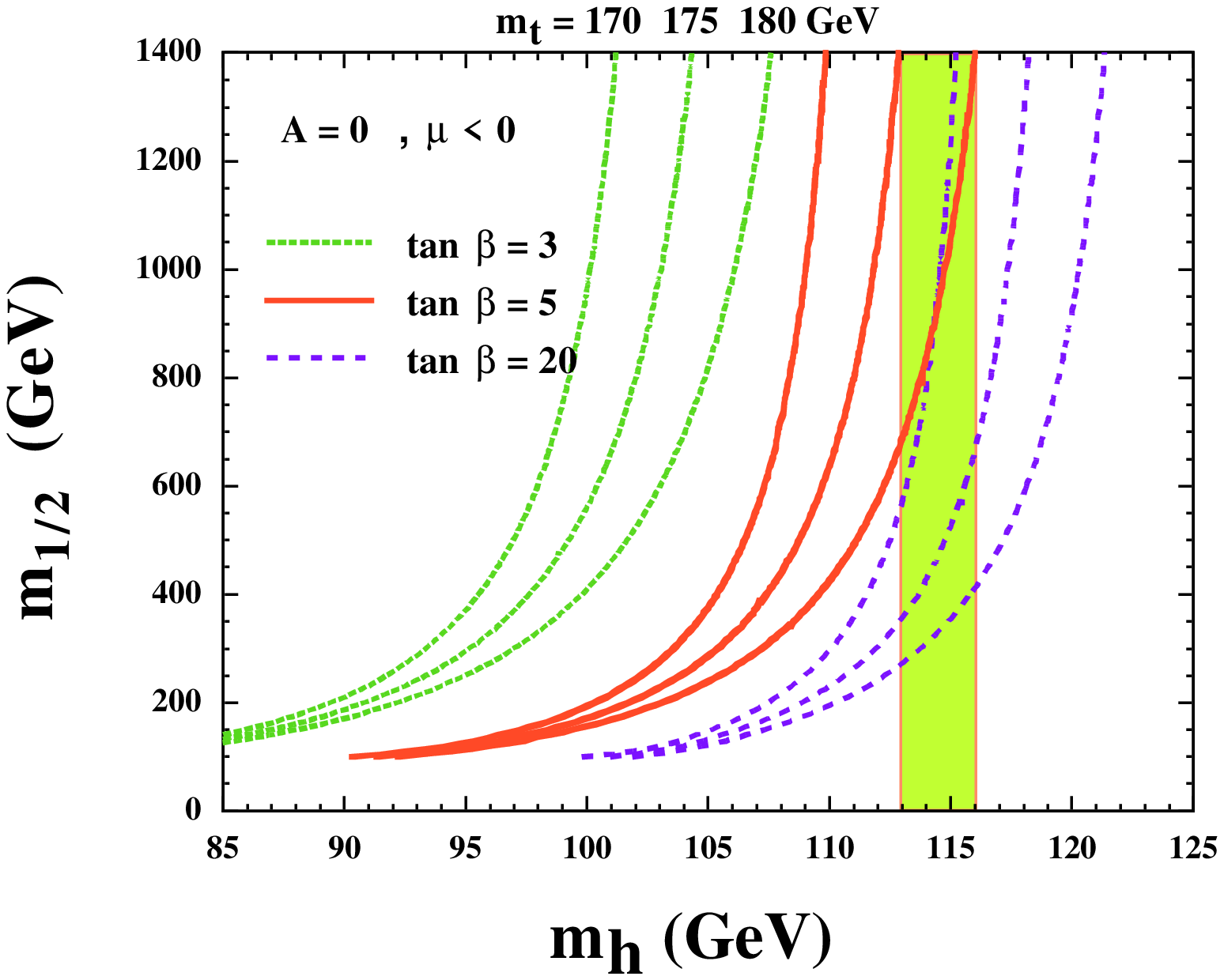,height=15cm}}
\end{center}
\vskip -1.3in
\caption[.]{\label{fig:A0}\it
The sensitivity of $m_h$ to $m_{1/2}$ in the CMSSM for (a) $\mu > 0$ and
(b) $\mu < 0$. The no-scale value $A = 0$ is assumed for definiteness.
The dotted (green), solid (red) and dashed (blue) lines are for
$\tan \beta = 3, 5$ and $20$, each for $m_t = 170, 175$ and $180$~GeV
(from left to right). The lines are relatively unchanged as one varies
$\tan \beta \ga 10$, where they are also insensitive to the sign of $\mu$.
The shaded vertical strip corresponds to $113~{\rm GeV} \le m_h
\le 116~{\rm GeV}$.
}
\label{fig:AO}
\end{figure}

The most important remaining uncertainty is that in $A$.  For
definiteness, henceforth we use as default value $A = 0$ at the input
scale, motivated
theoretically by no-scale supergravity models~\cite{noscale}, discussing
later the effect
of varying $A$ over a range of a few units in $m_{1/2}$. Panels (a) and
(b) of Fig.~\ref{fig:AO} show, for $\mu < 0 $ and $> 0$, contours of the
values of $m_{1/2}$ (vertical axis) required to obtain any given value of
$m_h$ (horizontal axis) for $\tan \beta = 3, 5$ and $20$ (from left to
right) and $m_t = 170, 175, 180$ GeV (also from left to right).
We have truncated the vertical axis at
$m_{1/2} = 1400$ GeV, which corresponds 
approximately to the maximum value of $m_{\chi}$
allowed by cosmology, which is attained when $m_{\chi} =
m_{\tilde{\tau}_R}$ for $\Omega_{\chi} h^2 = 0.3$, including
coannihilation effects~\cite{EFOSi,glp}. 
Since the curves for $\tan \beta = 10, 20$ are rather similar,
for clarity we do not plot any curves for $\tan \beta = 10$, nor for
$\tan \beta > 20$. We note also that
the high-$\tan \beta$ curves are relatively insensitive to the sign of
$\mu$.  On the other hand, the curves are quite different for smaller
$\tan \beta$, particularly $\tan \beta = 3$.  The vertical bands in
Fig.~\ref{fig:AO} correspond to $113~{\rm GeV} \le m_h \le 116~{\rm GeV}$,
including the `observed' range of $114$ to $115$ GeV, combined
with a theoretical error as discussed above. Requiring $m_h \ge 113$~GeV
clearly
imposes a non-trivial lower limit on $m_{1/2}$ and hence the sparticle
masses, as we discuss in more detail below.

\begin{figure}[htb]
\begin{center}
\vskip -3.0in
\mbox{\hskip -.7in \epsfig{file=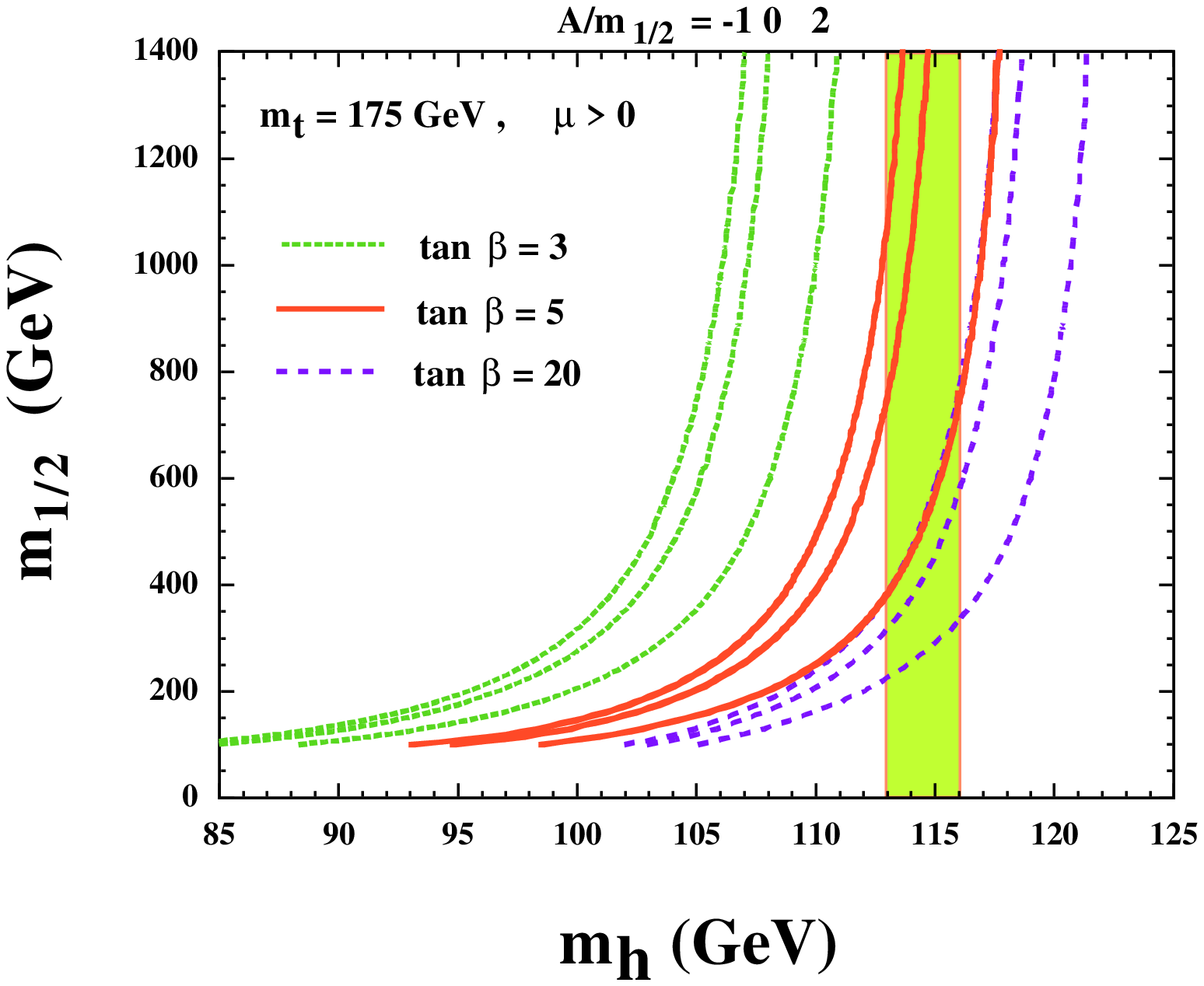,height=15cm}}
\vskip -2.1in
\mbox{\hskip -.7in \epsfig{file=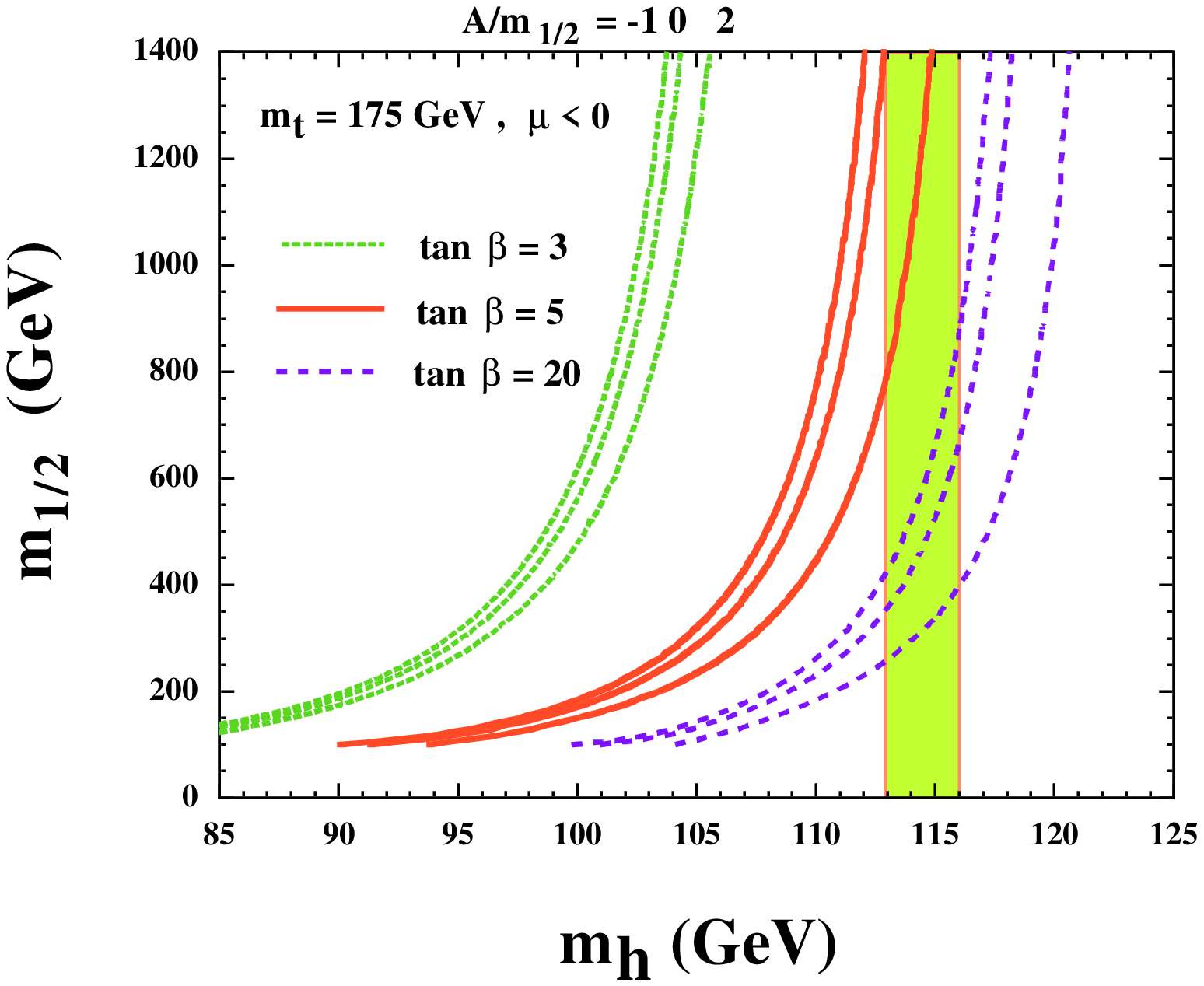,height=15cm}}
\end{center}
\vskip -1.3in
\caption[.]{\label{fig:varyA}\it
The sensitivity of $m_h$ to $m_{1/2}$ in the CMSSM for (a) $\mu > 0$ and 
(b) $\mu < 0$, this time showing the sensitivity to $A$, varied between
$ - m_{1/2}, 0$ and $+ 2 m_{1/2}$ (from left to right).
The dotted (green), solid (red) and dashed (blue) lines are again for
$\tan \beta = 3, 5$ and $20$, for $m_t = 175$~GeV.
The shaded vertical strip again corresponds to $113~{\rm GeV} \le
m_h
\le 116~{\rm GeV}$.
}
\end{figure}

Fig.~\ref{fig:varyA} illustrates the effect of varying $A$.  We see
that, for given values of $m_{1/2}, \tan \beta$ and the sign of $\mu$,
slightly lower values of $m_h$ are found for
$A = -m_{1/2}$ than for $A = 0$. Conversely, somewhat higher values of
$m_h$ are found for $A = + 2 m_{1/2}$~\footnote{We have also studied the 
case $A/m_{1/2} = 4$, for which the trend to higher $m_h$ continues. This
is near the maximum value of $A$ possible for $\tan
\beta = 3$ or 5, and is disallowed for $\tan \beta = 20$, because of
troubles with a light or tachyonic ${\tilde \tau}$~\cite{EFGO}.}
We note that the
differences in $m_h$ for these different values of $A$ and $m_t = 175$ GeV
are typically
less than those found by fixing $A$ and increasing (decreasing) $m_t$ to
$180 (170)$ GeV.  Therefore, in the following we restrict our attention to
the value $A = 0$ that we prefer on theoretical grounds~\cite{noscale},
varying $m_t$ between 170 and 180~GeV. 

It is apparent from Fig.~\ref{fig:efgo} that for $\tan \beta = 10$ the LEP
`value' of
$m_h$ pushes $m_{1/2}$ up into the $\chi - {\tilde \tau}$ coannihilation
region, which extends up to $m_{1/2} \sim 1400$~GeV~\cite{EFOSi}. The
necessity of including coannihilation effects is even more pronounced for
lower $\tan \beta$, since then the LEP `value' of
$m_h$ pushes $m_{1/2}$ even higher. The $m_h$ constraint is more
relaxed for larger $\tan \beta$, but then, for $\mu < 0$, the $b
\rightarrow s \gamma$ constraint also pushes $m_{1/2}$ into the
coannihilation
region~\footnote{There has recently been a suggestion~\cite{deBoer} that
the
$b \rightarrow s \gamma$ constraint at large $\tan \beta$ may be more
important for $\mu > 0$, but this is not supported by a recent NLO
analysis~\cite{newGambino}.}.

\begin{figure}[htb]
\begin{center}
\vskip -3.0in
\mbox{\hskip -.2in \epsfig{file=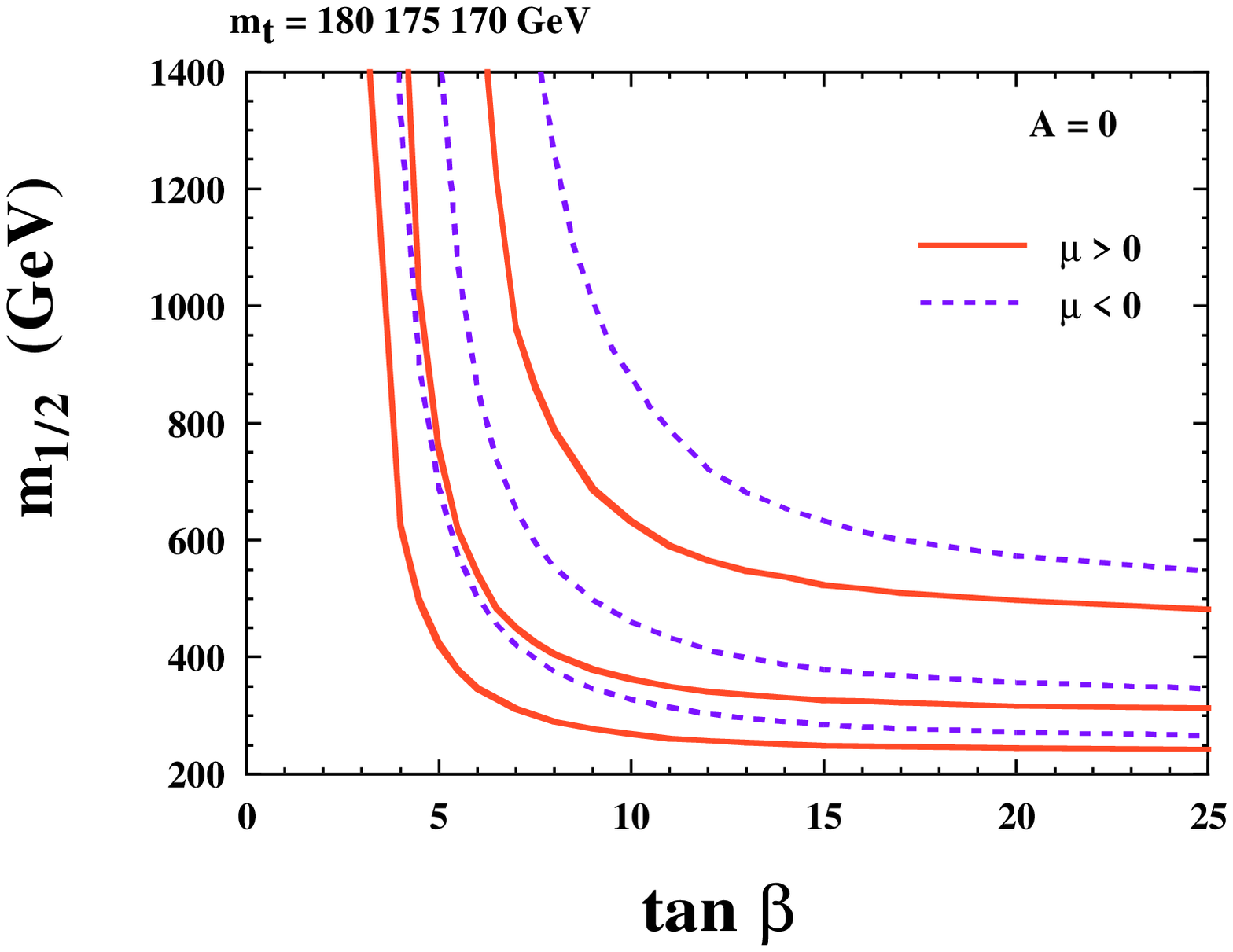,height=15cm}}
\vskip -2.1in
\mbox{\hskip -.2in \epsfig{file=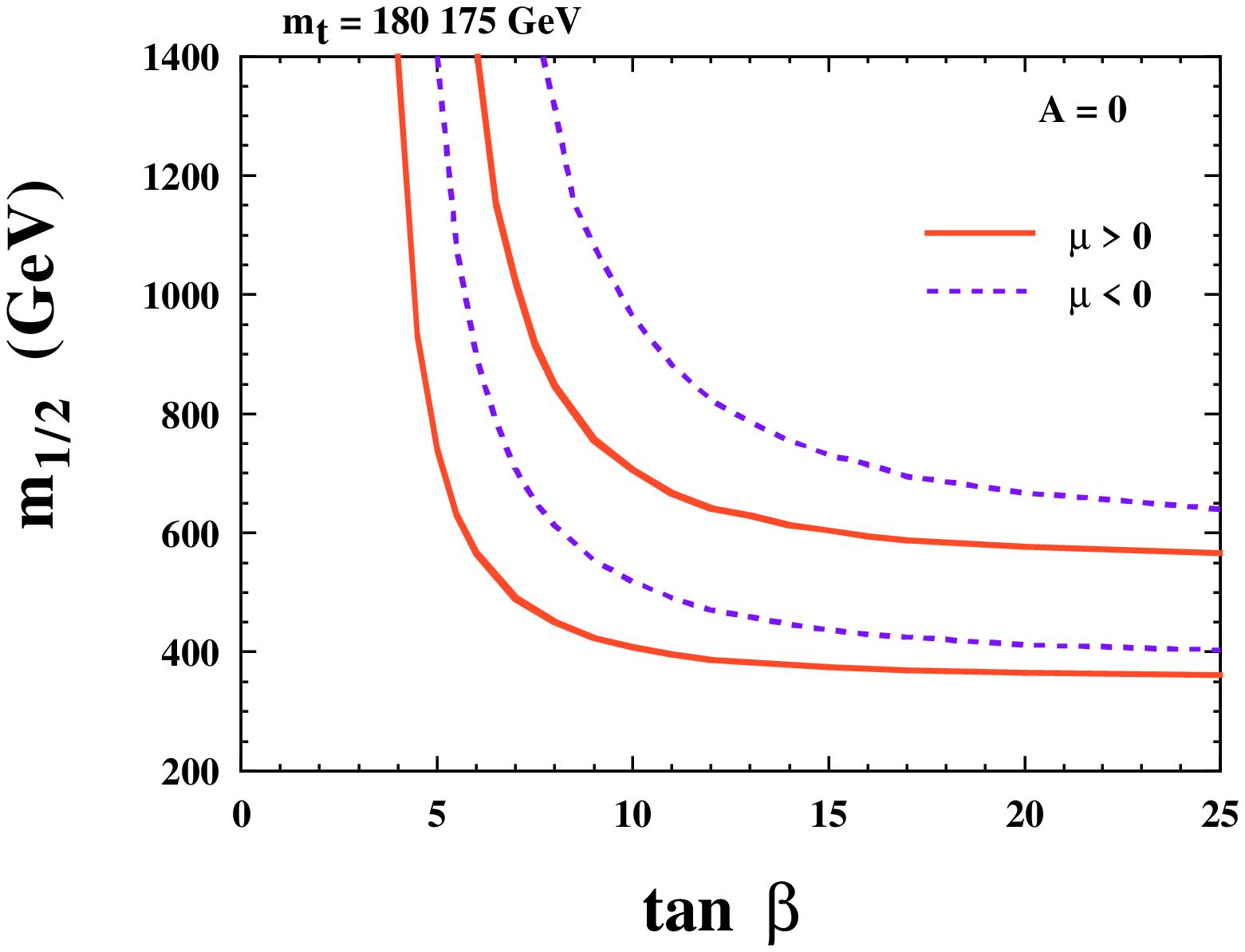,height=15cm}}
\end{center}
\vskip -1.3in
\caption[.]{
\it
(a) The lower limit on $m_{1/2}$ required to obtain $m_h \ge 113$~GeV for
$\mu > 0$ (solid, red lines) and $\mu < 0$ (dashed, blue lines), and $m_t
= 170, 175$ and $180$~GeV, and
(b) the upper limit on $m_{1/2}$ required to obtain $m_h \le 116$~GeV for
both signs of $\mu$ and $m_t = 175$ and $180$~GeV: if $m_t = 170$~GeV,
$m_{1/2}$ may be as large as the cosmological upper limit $\sim 1400$~GeV.
The corresponding values of the lightest neutralino mass $m_\chi \simeq
0.4 \times m_{1/2}$.
}
\label{fig:lowm12}
\end{figure}

We show in Fig.~\ref{fig:lowm12}(a) the lower bounds on $m_{1/2}$ obtained
assuming $m_h \geq 113$ GeV, for $\mu > 0$ (solid, red lines) and $\mu
< 0$ (dashed, blue lines), and $m_t = 170, 175$
and $180$ GeV (from bottom to top).  We note immediately a lower bound
\beq
m_{1/2} \gappeq 240 \; {\rm GeV},
\label{0}
\eeq
corresponding to a lower limit on the lightest neutralino $\chi$ of
\beq
m_{\chi} \gappeq 95 \; {\rm GeV},
\label{1}
\eeq
which is saturated for $\mu > 0$, $\tan \beta \sim 30$, and $m_t = 180$
GeV.  With the nominal value $m_t = 175$ GeV we would obtain $m_{1/2}
\gappeq 310$ GeV ($m_\chi \gappeq 125$~GeV). The lower bound on $m_{1/2}$
is not very sensitive to the sign
of $\mu$, particulary at large $\tan \beta$ as can be discerned from
Fig.~\ref{fig:lowm12}(a). On the other hand, the lower bound on
$m_{1/2}$ rises steeply for $\tan \beta \lappeq 10$, where it depends
more on the sign of
$\mu$. Recalling that
$m_{1/2} \sim 1400$ GeV is the maximum value of $m_{\chi}$
allowed by cosmology~\cite{EFOSi,glp}, we infer a lower bound
\beq
\tan \beta \gappeq 3
\label{2}
\eeq
attained again for $\mu > 0$ and $m_t = 180$ GeV.  The corresponding lower
limit for the nominal $m_t = 175$ GeV would be $\tan \beta \gappeq 4$.
For $\mu < 0$, the correspong limits are $\tan \beta \gappeq 4~(5)$ for 
$m_t = 180~(175)$. 

As we see in Fig.~\ref{fig:lowm12}(b), it is also possible in some cases
to derive an upper bound on $m_{1/2}$, obtained by requiring $m_h \leq
116$~GeV. The upper bound is relatively insensitive to the sign of
$\mu$ at large $\tan \beta$, and more sensitive at lower $\tan \beta$.
However, its greatest sensitivity is to the value of $m_t$, as seen in
Fig.~\ref{fig:lowm12}(b) for the cases $m_t = 175$ and $180$~GeV:
the corresponding maximum values of $m_{1/2}$ are $\sim 650$ and 400~GeV
for large $\tan\beta$, respectively. If
$m_t = 170$~GeV, the upper limit on $m_{1/2}$ in fact exceeds the upper
value $\sim 1400$~GeV allowed by the cosmological relic density, so this
case is not shown in Fig.~\ref{fig:lowm12}(b).

The
Tevatron collider may well be able to confirm at the 3-$\sigma$ level or
refute the LEP `observation'
of a Higgs boson with about $3 fb^{-1}$ of luminosity in each of CDF and
D0~\cite{Tevatron}. On the other hand,
the lower bound (\ref{0}) does not offer much encouragement for
$\tilde{q}$ and
$\tilde{g}$ searches at the Tevatron collider~\cite{sTevatron}, since one
expects
\begin{equation}
m_{\tilde{q}} \gappeq 600~{\rm GeV}, \; \;
m_{\tilde{g}} \gappeq 700~{\rm GeV}
\label{strongmasses}
\end{equation}
for $m_{1/2} \gappeq 240$~GeV~\footnote{The third-generation squarks
might be somewhat lighter, because of mixing.}. The
search for associated production of
charginos and neutralinos may offer brighter prospects~\cite{Barger},
but a definite conclusion
on this would require a more detailed study than is currently available. 
Examples of the estimated Tevatron sensitivity in this channel are shown
in Fig.~\ref{fig:efgo}~\cite{sTevatron}. We see that, in these particular
cases, the
chargino/neutralino process is also expected to be
unobservable~\footnote{Less glamorously, an improved measurement of $m_t$ 
from the Tevatron would be a significant contribution to pinning down
the interpretation of the LEP Higgs `signal'.}.
However, ATLAS and CMS at the
LHC should be able to detect both the Higgs boson and sparticles with high
significance~\cite{LHC}.

A Higgs boson weighing 114 to 115~GeV would be a bonanza for a sub-TeV
linear
$e^+e^-$ collider (LC), which would produce it copiously and study its
properties in detail~\cite{LC}.  However, its prospects for detecting
supersymmetry
would depend on the threshold for producing sparticle pairs~\cite{EGO},
for which the best prospects may be sleptons:
\begin{equation}
m_{{\tilde \ell}_R}^2 \simeq m_0^2 + 0.15 m_{1/2}^2, \; \;
m_{{\tilde \ell}_L}^2 \simeq m_0^2 + 0.52 m_{1/2}^2.
\label{smasses}
\end{equation}
and charginos $\chi^\pm$. 
The upper bound on $m_{1/2}$ imposed by the cold dark matter constraint
$\Omega_{\chi} h^2 \leq 0.3$ was used previously to estimate the maximum
energy required by a LC to be sure of seeing supersymmetry, namely $E_{cm}
\simeq 1.25$ TeV~\cite{EGO}. 
Looking back
at Fig.~\ref{fig:efgo}, in the context of our analysis we see that the
conservative way to bound
sparticle production from above
is first to take the lowest possible value of $m_{1/2}$. Then one
should choose the lowest 
value of $m_0$ for this value of
$m_{1/2}$, allowing the relic density to fall below $\Omega_{\chi} h^2 =
0.1$, as could occur if there is another source of cold dark matter.
Generically, this absolute lower bound $m_0^{\rm min}$ is found by
requiring
$m_{\chi} \geq m_{\tilde{\tau}_r}$, so as to avoid charged dark matter,
but in some cases this is not a constraint, and $m_0 = 0$ is allowed.

\begin{figure}[htbp]
\begin{center}
\vskip -0.5in
\mbox{ \epsfig{file=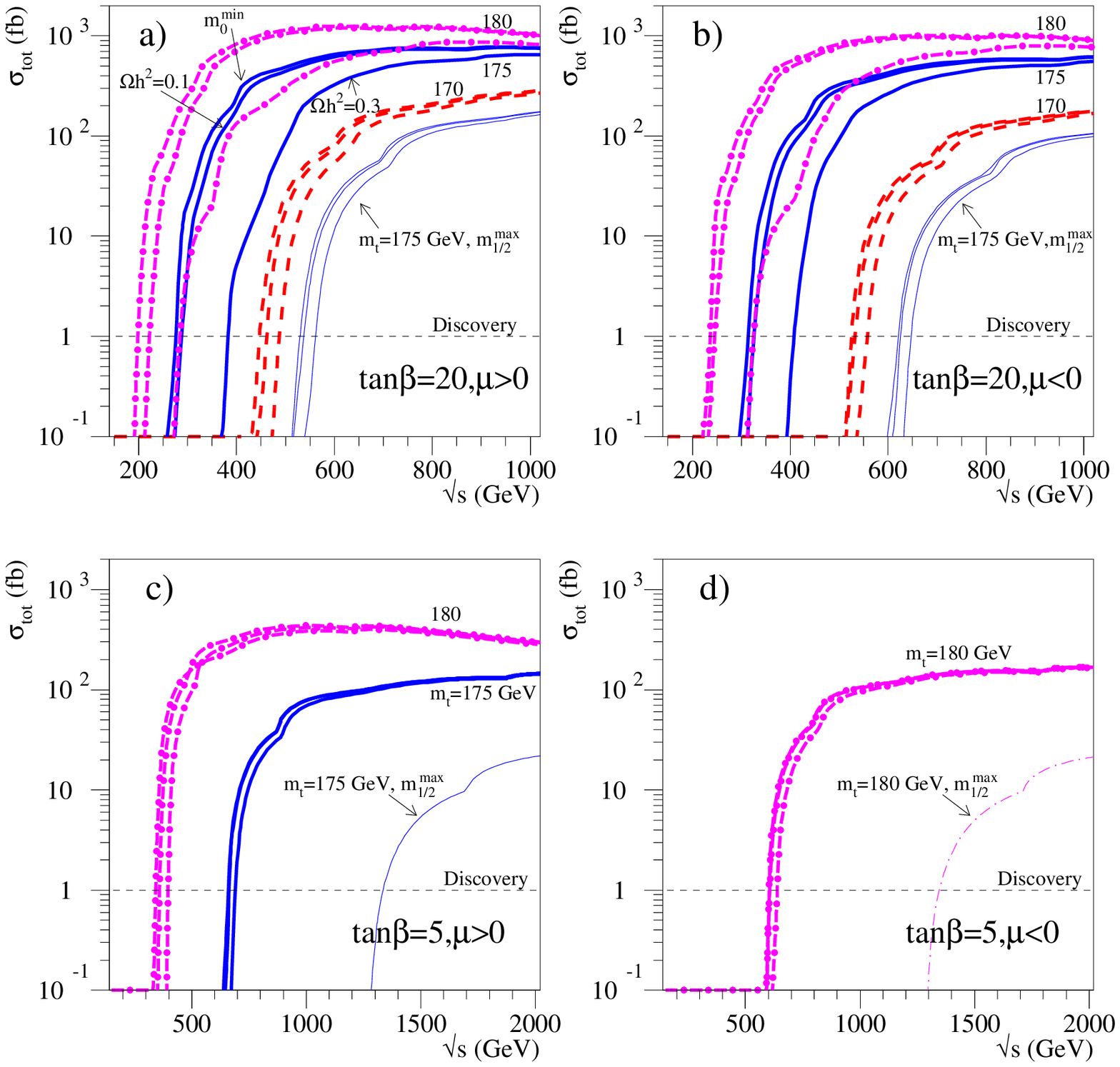,height=18cm}}
\end{center}
\caption[.]{
\it Cross sections for sparticle pair production at a linear $e^+ e^-$
collider, for (a) $\tan \beta = 20, \mu > 0$, (b) $\tan \beta = 20, \mu
< 0$, (c) $\tan \beta = 5, \mu > 0$ and (d) $\tan \beta = 5, \mu < 0$, as
functions of the centre-of-mass energy $\sqrt{s}$, compared with a
nominal discovery limit~\cite{EGO}.
The dashed (red), solid (blue) and dot-dashed (pink) lines are for
$m_t = 170, 175$ and $180$~GeV, respectively. The thicker (thinner)
lines are for the minimum (maximum) values of $m_{1/2}$. The different
lines in each style correspond to different choices of $m_0$: those
leading to $\Omega_\chi h^2 = 0.3$ and $0.1$, and the lowest allowed
value, disregarding the value of the relic density. In panels (c) and
(d), the maximum $m_{1/2} \simeq 1400$~GeV is taken, for which there is
only one allowed value of $m_0$.
}
\label{fig:Ethr}
\end{figure}

We show in Fig.~\ref{fig:Ethr} as thick lines the conservative upper
limits on the sum of the production cross sections for ${\tilde
\ell}^+ {\tilde \ell}^-$ and chargino pairs
found in this way. The kinks in the curves reflect the different
${\tilde \ell}_{L,R}$ and $\chi^\pm_i$ thresholds. There are
in general three lines for each 
choice of $\tan \beta$, the sign of $\mu$ and $m_t$, corresponding to
the values of $m_0$ that yield $\Omega_\chi h^2 = 0.3$ and $0.1$, and
the lowest value $m_0^{\rm min}$, disregarding the relic density. The
latter generally
gives the largest cross sections of all. We see in panel (a) of
Fig.~\ref{fig:Ethr}, for $\tan \beta = 20$ and $\mu > 0$, that
in this case a LC with $\sqrt{s} = 500$~GeV would be well placed to
discover supersymmetry, {\it if $m_{1/2}$ is close to its minimum
value}, whatever the value of $m_t$. Although panel (b) for $\tan \beta =
20$ and $\mu < 0$ is qualitatively similar, we see that in this case
the discovery of supersymmetry might be possible {\it only if
$m_t \ge 175$~GeV}. Panel (c) for $\tan \beta = 5$ and $\mu > 0$ is an
example where the discovery of supersymmetry might be possible 
with a $\sqrt{s} = 500$~GeV LC only if $m_t = 180$~GeV, and panel (c) for
$\tan \beta = 5$ and $\mu < 0$ is an example where discovery would not be
possible for any of the values of $m_t$ studied.

The thinner lines in Fig.~\ref{fig:Ethr} correspond to the maximum values
of $m_{1/2}$ discussed earlier, corresponding to $m_h \la 116$~GeV. In
panels (a) and (b) for $\tan \beta = 20$, only the case
$m_t = 175$~GeV is
shown: the thresholds for $m_t = 170$~GeV are beyond $\sqrt{s} =
1200$~GeV, and those for $m_t = 180$~GeV are similar to the curves
for minimal $m_{1/2}$ and $m_t = 175$~GeV. Discovery of supersymmetry
with a $\sqrt{s} = 500$~GeV LC could be `guaranteed' only if $m_t =
180$~GeV, not for $m_t \le 175$~GeV. In panels (c) and (d) for $\tan \beta
= 5$,
we see that the discovery of supersymmetry cannot be `guaranteed' for any
value of $m_t$. In panel (c), the cross sections for the maximum value
of $m_{1/2}$ when $m_t = 175$~GeV are similar to those for the minimum
value of $m_{1/2}$ when $m_t = 180$~GeV.

This analysis is not conclusive, but it does suggest that a linear
$e^+ e^-$ collider with $\sqrt{s} = 500$~GeV has a chance of discovering
supersymmetric particles. However, its prospects depend on unknowns
such as $m_t, \tan \beta$ and the sign of $\mu$, and the `measurement'
of $m_h$ does not guarantee success.

We now discuss the impact of combining the `observed'
value of $m_h$ with the measured rate~\cite{bsgamma} for $b \to s \gamma$
decay~\cite{Gambino}. 
For either sign of $\mu$, the $b \to s \gamma$
constraint excludes a region of the $(m_{1/2}, m_0)$ plane that extends to
larger $m_{1/2}$ as $\tan \beta$ increases, as exemplified in
Fig.~\ref{fig:efgo}(b) for $\tan \beta = 10, \mu < 0$.  On the
other hand, the value
of $m_{1/2}$ required to allow $m_h \le 116$ GeV decreases as $\tan \beta$
increases. Comparing the two constraints, we find for $m_t = 175$~GeV
that
\beq
\tan \beta \lappeq 25
\label{3}
\eeq
for $\mu < 0$, and for $m_t = 180$~GeV that $\tan \beta \lappeq 13 (33)$
for $\mu < 0 (>0)$. On the other hand, there is no reasonable upper
limit on $\tan\beta$ for $m_t = 170$~GeV, or for $m_t = 175$~GeV and
$\mu > 0$, since the upper bound imposed on $m_{1/2}$ is beyond the
reach of the constraints from $b \to s \gamma$. 

We comment finally on the prospects for direct detection of cold dark
matter by elastic scattering, within the CMSSM. The lower limit on
the lightest neutralino mass suggested by our analysis is $m_\chi \ga
95$~GeV. This is considerably stronger than was quoted in~\cite{EFGO},
essentially for two reasons. One is that the sensitivity of the LEP
experiments to MSSM Higgs bosons has exceeded our prognostications.
More significantly, here we estimate the $m_h$ sensitivity of the LEP
experiments
by calculating for each CMSSM parameter choice the corresponding $ZZh$
coupling strength, whereas previously we (too) conservatively used the
prospective LEP limits based on the maximal mixing scenario~\cite{MAX}.
In this scenario, the $ZH$ production cross section may be suppressed by a
factor $\sin^2(\alpha - \beta) \ll 1$, which we do not find in the CMSSM.

The strengthened lower limit on $m_\chi$ has the immediate effect of
decreasing the maximum elastic scattering cross section attainable in
the CMSSM~\cite{EFO}, from $\sim 10^{-4}$~pb to $\sim 10^{-5}$~pb in the
spin-dependent case
and $\sim 10^{-7}$~pb to $\sim 10^{-8}$~pb in the spin-independent case.
However, we emphasize that these upper limits apply for $\tan\beta \le
10$. For larger $\tan \beta$, the scalar
elastic scattering cross sections may be an order of magnitude larger
\cite{arnnop} (though we note that the scalar cross section is most
sensitive to 
$\tan \beta$ for $\mu < 0$ where the constraints from $b \to s \gamma$
are most restricitive). Larger cross sections may also be obtained if our
CMSSM assumption of universal scalar masses at the GUT scale is relaxed
\cite{efo2}.

We have shown in this paper how a measurement of the mass of the Higgs
boson may provide much valuable information, at least in a particular
theoretical context.  We re-emphasize that there may well not be a Higgs
boson weighing around 115 GeV, that supersymmetry may not exist, that
our model-dependent assumptions within the MSSM may be unjustified, that
the cold dark matter may not consist of neutralinos, etc.  Nevertheless,
we hope this paper serves a useful purpose in helping to focus attention
on ways in which any Higgs signal might be corroborated by other
experiments, in particular those looking for sparticle production at
colliders. Even if we must wait several years for the truth about a
possible Higgs boson weighing around 115 GeV to emerge, experiments at
the
Tevatron and elsewhere may aid in the interpretation of the possible
`signal'.

\vskip 0.5in
\vbox{
\noindent{ {\bf Acknowledgments} } \\
\noindent  
We would like to thank P. Gambino and especially
T. Falk for useful discussions.
The work of D.V.N. was partially supported by DOE grant
DE-F-G03-95-ER-40917, and that of
K.A.O. by DOE grant DE--FG02--94ER--40823.}

\end{document}